\documentclass[conference]{IEEEtran}
\usepackage{cite}
\usepackage{amsmath,amssymb,amsfonts}
\usepackage{algorithmic}
\usepackage{graphicx}
\usepackage{textcomp}
\usepackage{xcolor}
\usepackage{array}
\usepackage{mdwmath}
\usepackage{mdwtab}
\usepackage{eqparbox}
\usepackage{booktabs}
\usepackage{commath}
\usepackage{color}
\usepackage{bm}
\usepackage{hyperref}
\usepackage{tocloft}
\begin{document}

\title{Safe Reinforcement Learning-based Automatic Generation Control}

\author{\IEEEauthorblockN{Amr S. Mohamed, Emily Nguyen, and Deepa Kundur}
\IEEEauthorblockA{The Edward S. Rogers Sr. Department of Electrical \& Computer Engineering, University of Toronto\\
Toronto, Ontario, Canada, M5S 3G4\\
\{\texttt{\{amr.mohamed, emily.nguyen\}@mail., dkundur@}\}\texttt{utoronto.ca}
}
}

\maketitle

\begin{abstract}
Amidst the growing demand for implementing advanced control and decision-making algorithms|to enhance the reliability, resilience, and stability of power systems|arises a crucial concern regarding the safety of employing machine learning techniques. While these methods can be applied to derive more optimal control decisions, they often lack safety assurances.
This paper proposes a framework based on control barrier functions to facilitate safe learning and deployment of reinforcement learning agents for power system control applications, specifically in the context of automatic generation control. We develop the safety barriers and reinforcement learning framework necessary to establish trust in reinforcement learning as a safe option for automatic generation control|as foundation for future detailed verification and application studies.
\end{abstract}


\section{Introduction} \label{sec:introduction}

Grid optimization is a fundamental area of research in power systems, driven by the increasing complexity of the electric grid and the growing demand for enhanced efficiency and reliability. The continuous advancement of grid optimization methods is essential to meet evolving challenges to power system operation.
Recent advancement in machine learning has propelled grid optimization research~\cite{ibrahim2020machine}. One prominent machine learning approach is Reinforcement Learning (RL), which involves training an agent to learn optimal actions through iterative trial-and-error interactions. 
In the field of power systems, RL has found application in various domains, including demand response, operational control, economic dispatch, congestion management~\cite{zhang2019deep, glavic2017reinforcement}, as well as cybersecurity~\cite{zhang2019deep, mohamed2023use}.

A key advantage of RL is its adaptability to dynamic, uncertain environments, making it ideal for the evolving energy landscape, including load shifts from demand response, electrification, and renewable integration. It excels in optimizing decisions for complex, non-linear systems that are difficult to model with traditional approaches.

% A distinctive advantage of RL lies in its adaptability to dynamic and uncertain environments; a critical advantage in the face of the rapidly changing energy landscape. This includes adapting to shifts in load dynamics, such as due to demand response and electrification, as well as the integration of renewable energy sources. 
% RL is also well suited to optimizing decisions for non-linear and highly complex real system characteristics that are challenging to model for the development of traditional control and optimization approaches.

Despite its promise, a significant challenge remains in ensuring the safety of RL decisions. As RL agents learn through interaction, there is a potential for making unsafe decisions during learning or adapting. 
This concern is particularly pertinent in power systems, where safety is paramount and even minor deviations from optimal actions can lead to significant consequences. 
Consequently, there is a growing focus on the development of safe RL techniques, providing a framework to ensure that learned policies optimize operation while incorporating safety assurances to prevent undesirable outcomes. 
The field of safe RL in power systems is still in its early stages, with only a few papers, including~\cite{vu2021barrier, tabas2022computationally, cui2022decentralized}, incorporating safety in training RL. 

In this study, we extend the application of safe RL in power systems to Automatic Generation Control (AGC) persuaded by its pivotal role in power system stability. 
As a testament to its pivotal role, AGC has been the subject of extensive grid optimization research. Previous studies have sought to enhance AGC through various approaches, such as classical optimal PI control~\cite{saikia2011performance}, evolutionary~\cite{abdel2003agc, babu2020application, pradhan2016firefly} and genetic~\cite{panda2013automatic, demiroren2007ga} algorithms, %abdel1996optimal
fuzzy logic~\cite{pradhan2016firefly, chandrakala2013variable, arya2019new}, neural networks~\cite{karnavas2002agc}, and RL~\cite{ahamed2002reinforcement, li2020deep, liu2022agc, zhang2024safe}.
% chown1998design fuzzy
A significant research gap remains in developing safe RL algorithms for AGC.

RL stands to offer benefits for AGC dynamic optimization by employing a data-driven control strategy that integrates diverse system data, such as renewable energy forecasts, and enables dynamic responses to power disturbances~\cite{liu2022agc, zhang2024safe}. It reduces reliance on hard-to-obtain accurate system models~\cite{ahamed2002reinforcement, zhang2024safe} and supports complex security constraints~\cite{liu2022agc} and advanced regulation payment structures~\cite{li2020deep}. This paper focuses on a centralized RL-based AGC approach, where the RL agent supplements or replaces the centralized traditional AGC controller, with potential for future application of multi-agent RL-based decentralized AGC systems.

% In line with existing literature, RL stands to offer benefits for the dynamic optimization of AGC by employing a data-driven control strategy that enables the integration of diverse system data~\cite{liu2022agc, zhang2024safe}, including an increasing array of energy sources and renewable energy forecasts, and allows for dynamic responses to continuous power disturbances~\cite{liu2022agc}. Additionally, RL mitigates the reliance on precise system models, which are often challenging to obtain~\cite{ahamed2002reinforcement, zhang2024safe}. RL has also been identified as a promising approach for incorporating complex security constraints~\cite{liu2022agc} and advanced regulation payment structures for AGC participants~\cite{li2020deep}. While this paper focuses on a centralized RL-based AGC approach, where an RL agent either replaces or augments the traditional centralized AGC controller, multi-agent RL frameworks present opportunities for implementing distributed RL-based AGC systems.

In our approach to designing safe RL, we introduce a novel formulation of control barrier functions (CBFs) to guarantee the safety of AGC, effectively mitigating potentially unsafe RL actions. We empirically demonstrate that this CBF-based approach effectively prevents AGC failures during the learning process and enables the agent to converge to a highly efficient AGC policy.
To the best of our knowledge, this research involves the first application of safe RL with formal safety guarantees to AGC. 
Below, we formulate the barrier functions, develop an actor-critic deep RL agent, and empirically verify the safety of the method on a two-area linearized power system. We briefly discuss a plan for necessary future work in Section V.
% We anticipate that our work will serve as a compelling case for the wide adoption of safe RL in other power system applications, leveraging the novel concepts presented in this paper.

The paper is structured as follows: Section II provides a background on CBFs. In Section III, we outline the methodology for integrating CBFs and RL within a safe RL framework for AGC. Results are presented in Section IV, followed by conclusions and directions for future research in Section V.

\section{Background}

CBFs seek to establish a safety set, which represents the states in which the system is considered safe, and enforce the forward invariance of this safety set, meaning that the system is forced to remain within this set for all future times. 

In control applications, CBF permits only control actions that maintain the system within the defined safety set. When integrated into the learning process of RL agents, CBF can ensure that the agent's actions adhere to safety constraints, thereby promoting a safe learning process.
The safe set is defined by the super-level set of a continuously differentiable function $h : \mathbb{R}^n \rightarrow \mathbb{R}$:

\begin{align}
    \mathcal{S} &= \{x \in \mathbb{R}^n : h(x) \geq 0 \} \nonumber\\
    \partial \mathcal{S} &= \{x \in \mathbb{R}^n : h(x) = 0 \}\\
    \text{Int}(\mathcal{S}) &= \{x \in \mathbb{R}^n : h(x) > 0 \} \nonumber
\end{align}

In this paper, we employ logarithmic barrier functions, inspired by the work in \cite{ames2016control, mohamed2023safety}. 
The function $\mathcal{B}(x)$ is considered a logarithmic barrier function if there exists a strictly increasing function $\alpha: \mathbb{R} \rightarrow \mathbb{R}$ with $\alpha(0) = 0$, such that for the control system $\dot{x} = f(x) + g(x)u$, the function $\mathcal{B}(x)$ satisfies the inequality:
\begin{align} \label{eq:cbf_log}
\dot{\mathcal{B}} = \frac{\partial \mathcal{B}}{\partial x} \dot{x} = \frac{\partial \mathcal{B}}{\partial x} \left( f(x) + g(x)u \right) \leq \alpha \left( \frac{1}{\mathcal{B}(x)} \right)
\end{align}

The logarithmic barrier function $\mathcal{B}(x)$ is computed based on $h$ as follows:
\begin{equation}
    \mathcal{B}(x) = -\log \left( \frac{h(\bm{x})}{1+h(\bm{x})} \right)
\end{equation}

and exhibits two characteristics:
\begin{align}
    \inf_{x \in \mathcal{S}} \mathcal{B}(x) &> 0\\
    \lim_{x \rightarrow \partial \mathcal{S}} \mathcal{B}(x) &= \infty
\end{align}
i.e., 
(1) Within the safety set, the logarithmic barrier function is positive, with no discontinuities. 
When the system is safe, $x$ is in the interior of $\mathcal{S}$, and consequently, $\mathcal{B}(x)$ and $\alpha(1/\mathcal{B}(x))$ are positive. This allows $\mathcal{B}(x)$ to increase toward the system boundary, if needed, while still remaining in the safety set.
The logarithmic barrier function is asymptotic near the boundary of the safety set, i.e., $\mathcal{B}(x) \rightarrow \infty$, enforcing a soft constraint, allowing the system to approach the safety boundary without violating it.
(2) At the boundary, $\alpha(1/\mathcal{B}(x)) = 0$, forcing $\mathcal{B}(x)$ to either stay at the boundary or decrease, moving away from the boundary into the interior of the safety set. 
Proofs are provided in \cite{ames2016control}.

Within the context of safe RL, CBFs can avert unsafe actions:
\begin{enumerate}
    \item by signaling when the control input~$u$ is expected to lead the system into unsafe regions, as indicated by the failure of condition~\eqref{eq:cbf_log}; or
    \item by adjusting the control input~$u$ to maintain system safety. This can be achieved by solving the following optimization problem:
    \begin{align} \label{eq:sccOpt}
        u = \arg & \min_{\Tilde{u}} & \frac{1}{2} \|u - \Tilde{u} \|_2\\
        & \text{s.t.} & \frac{\partial \mathcal{B}}{\partial x} \left( f(x) + g(x)\Tilde{u} \right) \leq \alpha \left( \frac{1}{\mathcal{B}(x)} \right)\nonumber 
    \end{align}
\end{enumerate}

\section{Method}

\subsection{System Model}

Defining a CBF to guarantee the safety of RL-based AGC requires a system model that encompasses the load frequency dynamics of the power system. 
The state-space system, represented by~\eqref{eq:state-space}, serves as a general model for the load frequency dynamics of a power system. In this model, the state vector $\bm{x}$ encompasses the internal states of the generators, frequencies of the power system areas, and the tie-line power flows. The input vector $\bm{P}_L$ accounts for the power demands in each area, while the input vector $\bm{P}_{ref}$ comprises the AGC reference power signals for all AGC-participating generators. The output vector $\bm{f}$ extracts the frequencies of the power system areas.
% The process of deriving this state-space representation is a well-established procedure. 
For a detailed explanation on how to construct such a model for a power system, we direct interested readers to~\cite{kundur2007power}.
\begin{align}\label{eq:state-space}
    \dot{\bm{x}} &= \bm{A} \bm{x} + \bm{B}_1 \bm{P}_L +  \bm{B}_2 \bm{P}_{ref}\\
    \bm{f} &= \bm{C} \bm{x}\nonumber
\end{align}

% \begin{align} \label{eq:state-space}
% 	\dot{\bm{x}} &= \bm{A} \bm{x} + \bm{B}_L \bm{P}_L +  \bm{B}_R \bm{P}_{ref}\\
%     &= \bm{A} \bm{x} + \bm{B}_L \begin{bmatrix} P_{L1} \\ P_{L2} \end{bmatrix} +  \bm{B}_R \begin{bmatrix} P_{ref,G1} \\ P_{ref,G2} \\ P_{ref,G3} \\ P_{ref,G4} \end{bmatrix} \nonumber\\
% 	\bm{y} &= \begin{bmatrix} f_{A1} \\ f_{A2} \end{bmatrix} = \begin{bmatrix} \bm{c}_{A1} \\ \bm{c}_{A2} \end{bmatrix} \bm{x} = \bm{C} \bm{x}
% \end{align}

\subsection{Control Barrier Function}

The CBF-based approach must issue prior warning to when the AGC signals, included in $\bm{P}_{ref}$, might be projected to violate the system's safety. 
A primary responsibility of AGC is to dispatch AGC-participating generators to stabilize frequency in response to power imbalances. 
The safety of the system may be compromised if generators are dispatched incorrectly, failing to stabilize the frequency, or if erroneous dispatches result in frequency deviations.
Consequently, safety can represented by thresholds on the frequencies of the power system areas, which if surpassed can lead to system instability. 
Therefore, the CBF must ensure that the RL-based AGC signals do not allow or induce significant frequency deviations in any of the power system areas.

In the discretized state-space model, the frequencies of the areas at time $t + T_s$ are expressed as:
\begin{equation}
    \bm{f}_{t+T_s}(\bm{x}) = \bm{C} (\overline{\bm{A}} \bm{x} + \overline{\bm{B}}_1 \bm{P}_L + \overline{\bm{B}}_2 \bm{P}_{ref} )
\end{equation}
where
\begin{align}
    \overline{\bm{A}} &= e^{\bm{A} T_s}\\
    \overline{\bm{B}}_1 &= \bm{A}^{-1} (\overline{\bm{A}} - \bm{I}) \bm{B}_1\\
    \overline{\bm{B}}_2 &= \bm{A}^{-1} (\overline{\bm{A}} - \bm{I}) \bm{B}_2
\end{align}

Using the discretized state-space model, the safety set $\mathcal{S}$ can be defined as follows:
\begin{align} \label{eq:safety}
	\mathcal{S} = \{& \bm{f}_{t+T_s} = [f_{i, t+T_s}]^{i \in \{1, \cdots, m\}} : \abs{f_{i, t+T_s}} \leq F \}
\end{align}
where $f_i$ is the frequency of area $i$, $F$ represent the maximum value of frequency deviation permitted in the power system areas, and $m$ is the number of areas.

Subsequently, the following CBF can be used to define the safety set in area $i$:
\begin{align} \label{eq:cbf_f}
    h_{i}(\bm{x}) &= F^2 - f_{i, t+T_s}^2(\bm{x})
\end{align}

% \begin{align} \label{eq:cbf_f1}
%     h_{A1}(\bm{x}) &= F^2 - f_{A1, t+T_s}^2(\bm{x})\\
%     \label{eq:cbf_f2}
%     h_{A2}(\bm{x}) &= F^2 - f_{A2, t+T_s}^2(\bm{x})
% \end{align}

The AGC signals do not cause violation of the system safety as long as in each area $i$
\begin{align}
    \Dot{\mathcal{B}}_{i}(\bm{x}) - \frac{\alpha_{i}}{\mathcal{B}_{i}(\bm{x})} &\leq 0
\end{align}

% \begin{align}
%     \Dot{\mathcal{B}}_{A1}(\bm{x}) - \frac{\alpha_{A1}}{\mathcal{B}_{A1}(\bm{x})} &\leq 0\\
%     \Dot{\mathcal{B}}_{A2}(\bm{x}) - \frac{\alpha_{A2}}{\mathcal{B}_{A2}(\bm{x})} &\leq 0
% \end{align}

where 
\begin{align}
    \Dot{\mathcal{B}}_{i}(\bm{x}) &= \frac{d \mathcal{B}_{i}(\bm{x})}{d \bm{x}} \Dot{\bm{x}} \nonumber\\
    &= - \frac{\frac{dh_{i}}{d\bm{x}}}{h_{i}(\bm{x}) + h_{i}^2(\bm{x})}(\bm{A} \bm{x} + \bm{B} \bm{P}_{ref})
\end{align}
\begin{align}
   \frac{dh_{i}}{d\bm{x}} &= -2 f_{i,t+T_s}(\bm{x}) \frac{d f_{i, t+T_s}}{d\bm{x}} \nonumber\\ 
    &= -2 f_{i,t+T_s}(\bm{x}) \bm{c}_{i} \overline{\bm{A}}
\end{align}
and $\bm{c}_{i}$ is row~$i$ of matrix $\bm{C}$ in~\eqref{eq:state-space}.

\subsection{Safe Reinforcement Learning}

The input measurements to AGC comprise the tie-line power flows between areas and area frequencies. Similarly, in the design of the RL agent, we configure the agent to observe these measurements. Using these observations, the RL agent calculates AGC power reference signals for AGC-participating generators, denoted as the input vector $P_{ref}$ in~\eqref{eq:state-space}. Before dispatching AGC signals to the generators, the CBF block scrutinizes them to assess whether they could lead to a safety violation in the system. If not, the AGC signals are permitted; otherwise, they are blocked.

During offline training, we conclude the current RL training episode following a CBF flag to block AGC signals. The RL agent is rewarded based on the following reward function:
\begin{align} \label{eq:rewardFunction}
    R &= -\bm{r}_1^T \abs{\bm{f}} - \bm{r}_2^T \abs{\bm{P}_{tie}} - r_3 \bm{P}_{ref}^T\bm{P}_{ref}\nonumber\\ 
    &- r_4 \sum_{i=1}^{m} \sum_{j \in \mathcal{G}_{i}} \left( P_{ref,i,j} - \frac{1}{|\mathcal{G}_{i}|} \sum_{k \in \mathcal{G}_{i}} P_{ref,i,k} \right)^2\\
    &- r_5 \{\text{CBF flag}\} - r_6 \{\text{Safety violation}\} \nonumber 
\end{align}

The first two terms penalize the agent proportionally to the deviation in area frequencies and tie-line power flows. To optimize AGC, the agent must minimize these deviations. 
The third term penalizes the agent based on the change of AGC power dispatch to encourage smaller dispatch changes. 
We introduce the fourth term to encourage equal load sharing between generators in the same area; this term can be easily modified to encourage another load sharing policy, e.g., proportional load sharing per generator capacities. Note that $\mathcal{G}_{i}$ denotes the set of generators in area $i$.
The last two terms impose significant penalties when the agent triggers a CBF flag or undermines the CBF, resulting in a safety violation.
Our inclusion of the last term is to demonstrates that the CBF prevents such events from occurring, serving as a safety guarantee to ensure RL actions do not violate safety. 
The hypothesis of the research is that the CBF-based approach prevents the agent from ever incurring the penalty expressed by the last term. We substantiate this hypothesis in the results.

After deployment (in an online setting), the CBF flag can alternatively be followed by reverting to a validated controller or modifying the RL actions using~\eqref{eq:cbf_f} to maintain safety while still receiving dispatch from RL-based AGC.

\begin{figure}
    \centering
    \includegraphics[width=0.9\columnwidth]{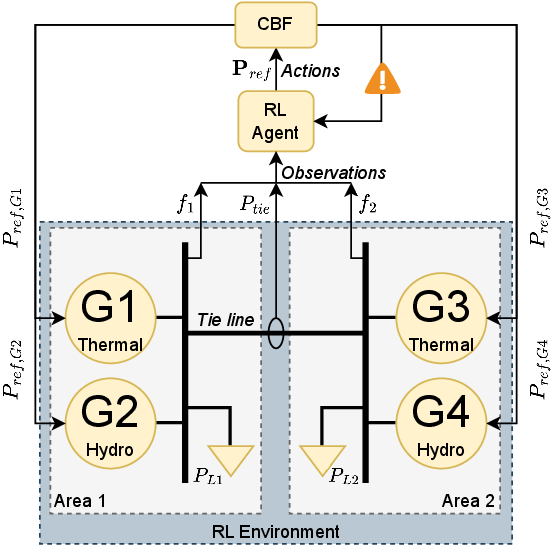}
    \caption{Illustration of CBF-based safe RL in a two-area power system.}
    \label{fig:safeRLblockdgm}
\end{figure}

This paper's focus is on developing CBFs to facilitate the safe learning of any RL agent and demonstrate their effectiveness in ensuring safety during learning. 
% This is to advocate for their use in safety-critical power system applications, including AGC. 
Therefore, we confine our discussion of the RL agent to the aforementioned details, complemented by the agent's neural network architecture in Fig.~\ref{fig:RLnetwork} in the appendix. For more details on developing deep RL agents for power system applications, interested readers can refer to~\cite{mohamed2023use}. 
Fig.~\ref{fig:safeRLblockdgm} illustrates the safe training of a RL agent for a two-area power system.

\section{Results}

We present our findings on a multi-source, multi-area system comprising two areas, each equipped with a thermal and a hydro generator. A tie line connects the two areas.
Each generator is equipped with a speed governor, functioning as the primary frequency controller to match power supply in the corresponding area with the demand. AGC serves as the secondary frequency controller, fine-tuning the areas' frequencies and tie line power.
The model, replicated from \cite{chandrakala2013variable}, is illustrated in Fig.~\ref{fig:safeRLblockdgm}.

In this section, we initially illustrate the purpose of employing CBF to explain their utilization in safe RL. Subsequently, we showcase that the incorporation of CBF effectively prevents safety violations during RL, all the while enabling the RL agent to learn a proficient AGC policy.

We use the following values in the generation of the results:
The RL agent takes actions in $2$ second intervals per typical timescale of AGC.
The CBF is designed with $\alpha_1 = \alpha_2 = 1$, $T_s = 0.5$ s, and $F = 0.4$ Hz.

\subsection{CBF-based safety} \label{subsec:results:cbf}

\begin{figure}
    \centering
    \includegraphics[width=0.95\columnwidth]{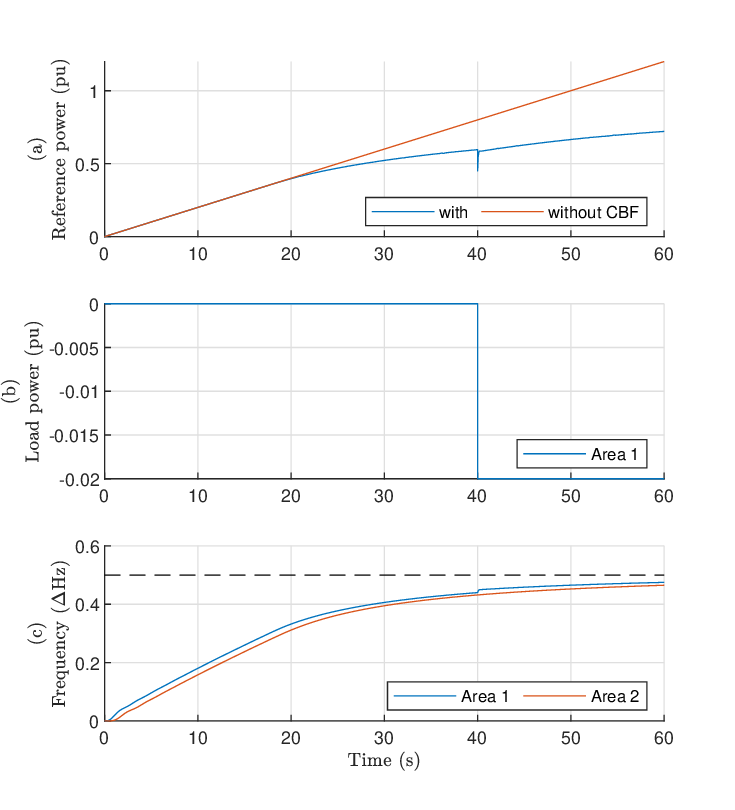}
    \caption{The CBF-based approach preemptively prevents erroneous dispatch from violating safety. This result is generated using $\alpha_1 = \alpha_2 = 0.1$, $T_s = 0.1$ s, and $F = 0.5$ Hz.}
    \label{fig:cbfdemo}
\end{figure}

An erroneous AGC reference power signal to generator G1 in Area 1 (refer to Fig.~\ref{fig:safeRLblockdgm}) is depicted by the blue curve in Fig.~\ref{fig:cbfdemo}(a). This erroneous AGC signal, if followed, would lead to a breach of the power system's safety set by causing an excess of power supply over demand, thereby inducing a frequency deviation in the two areas beyond $0.5$ Hz. 
A potential consequence could be the tripping of generators, resulting in system instability or cascading failure.
The CBF preempts the violation of safety by adjusting the AGC signal minimally, as illustrated by the leveling of the blue curve in Fig.~\ref{fig:cbfdemo}(a). The adjusted AGC signal is computed per~\eqref{eq:sccOpt}.

To further underscore the effectiveness of the CBF, at 40 seconds, a $2\%$ load change occurs in Area 1, as shown in Fig.~\ref{fig:cbfdemo}(b). The CBF reacts correspondingly to maintain system safety.
The area frequencies remain within the safety set, which is illustrated by the region beneath the dashed line in Fig.~\ref{fig:cbfdemo}(c).

Our safe RL approach leverages the CBF's ability to raise an alarm when signal modification is necessary to maintain safety. The alert is triggered when the red and blue curves in Fig.~\ref{fig:cbfdemo}(a) diverge.

\subsection{CBF-based safe RL}

\begin{figure}
    \centering
    \includegraphics[width=0.9\columnwidth]{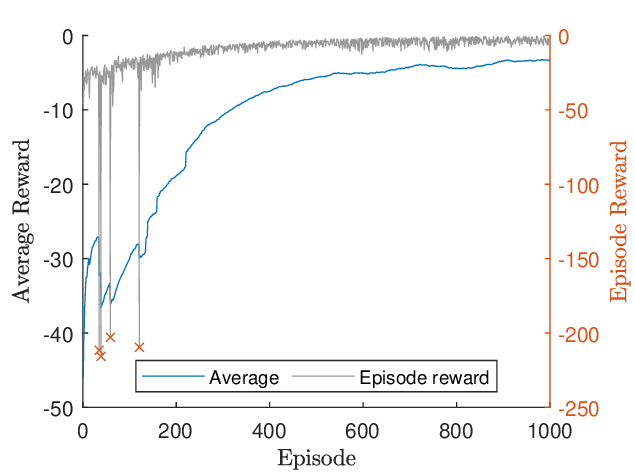}
    \caption{The CBF-based safe RL does not violate safety during training and converges to an effective AGC policy.}
    \label{fig:learnCurve}
\end{figure}

The coefficients used in training the RL agent in the reward function~\eqref{eq:rewardFunction} are $\bm{r}_1 = [2, 2]$, $r_2 = 40$, $r_3 = 25$, $r_4 = 15$, $r_5 = 200$, and $r_6 = 10^5$.

Fig.~\ref{fig:learnCurve} presents the episode and average rewards (over 100 episodes) obtained by the RL agent during training. The increasing and converging average reward indicates that the agent successfully learns an AGC policy. The episode rewards drop below $-200$ in a few episodes in the beginning, marked by the red crosses in Fig.~\ref{fig:learnCurve}, when the CBF alerts to the RL agent's unsafe actions. 
However, they never fall below $-10^5$, confirming the effectiveness of the CBF in preventing safety violation.

\subsection{Optimized AGC}

\begin{figure}
    \centering
    \includegraphics[width=\columnwidth]{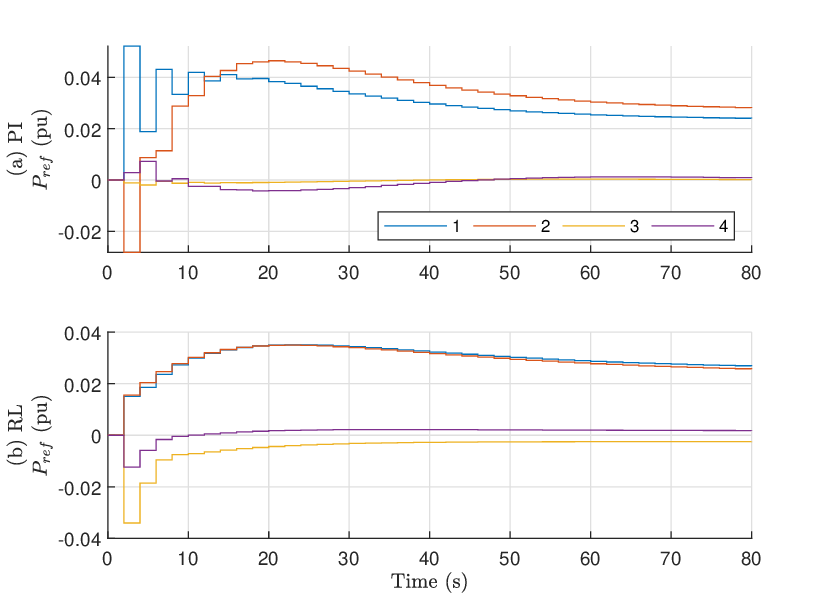}
    \caption{Comparing the AGC power reference signals dispatched by (a) PI control to (b) the RL agent.}
    \label{fig:inputCompare}
\end{figure}

\begin{figure}
    \centering
    \includegraphics[width=0.95\columnwidth]{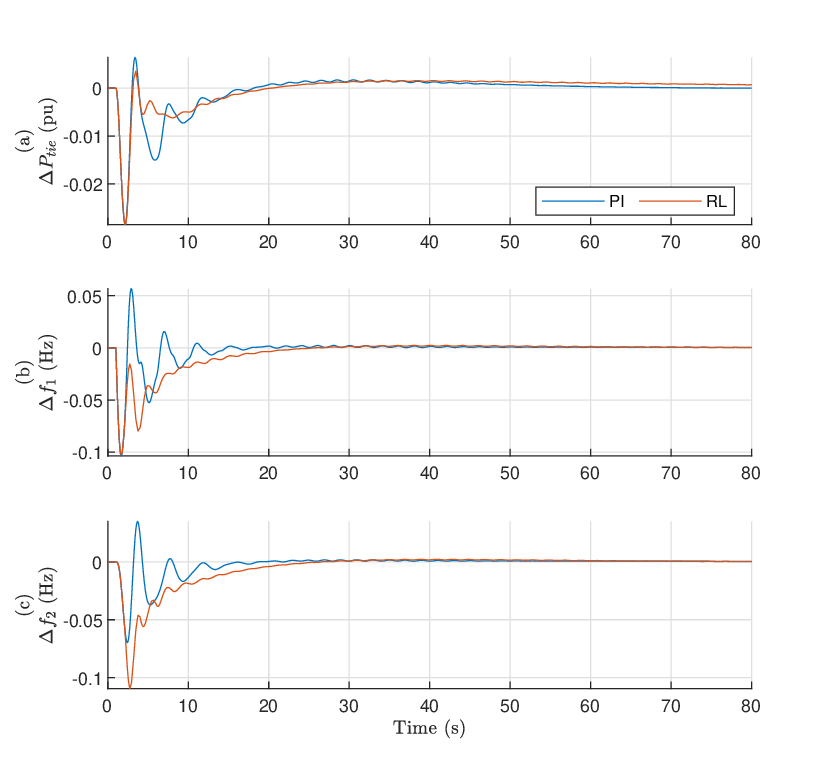}
    \caption{Comparing the AGC performance of PI control to the RL agent in terms of (a) tie-line power flow and (b) frequency deviations in Area 1 and (c) Area 2.}
    \label{fig:outputCompare}
\end{figure}

Finally, we present the AGC policy learned by the RL agent. 
We compare between the RL agent and a PI controller tuned using MATLAB's {PID Tuner} App. 
Fig.~\ref{fig:inputCompare} displays the reference power signals to the generators, while Fig.~\ref{fig:outputCompare} illustrates the tie-line power and frequency deviations in the areas. The AGC responds to a $5\%$ load change that occurs at time 1 second in Area 1.

The figures demonstrate that the RL agent effectively learns a control policy for AGC that ensures system stability and, in some aspects, outperforms the PI controller. Specifically, the RL agent's control leads to smaller dispatch changes and more balanced load sharing in Area 1, lower tie-line power deviation, and generally more damped system behavior, as shown in Fig.~\ref{fig:outputCompare}.

It is worth noting that further optimization of the RL agent's policy can be achieved through adjustments to the agent's reward function and hyperparameters. Our primary emphasis is on demonstrating the effectiveness of the CBF-based safe RL framework.

\section{Conclusion and Future Work}

% In light of the escalating demand for more intelligent control and decision-making algorithms to enhance power system reliability, resilience, and stability, there arises a crucial concern regarding the safety of deploying machine learning methods. While these methods excel at distilling vast amounts of power system data for optimized control decisions, they often lack safety guarantees.

In this paper, we developed a (safety) control barrier function-based framework for safe reinforcement learning (RL) in power system automatic generation control (AGC). We applied the framework to AGC in a multi-source, two-area power system. The results demonstrated that our framework ensures safety during online training, preventing and containing erroneous RL agent control decisions while allowing effective learning.

Future work will focus on practical implementation, addressing system non-linearity, deployment constraints, scalability for larger power systems, and further comparison with existing AGC methods.

% This paper presented the foundational work on formulating the safety barriers and empirically verifying their safety. Future work will expand the research to address practical implementation in real-world systems, tackling challenges like system non-linearity and physical deployment constraints, as well as discuss scaling the method with the scale of power systems, and comparing it to traditional AGC methods.
% \appendix
\begin{figure}
    \centering
    \includegraphics[width=0.55\columnwidth]{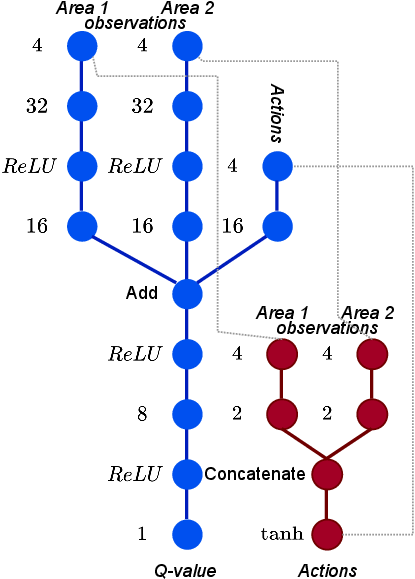}
    \caption{RL agent neural network architecture. The blue and red networks show the critic and actor, respectively.}
    \label{fig:RLnetwork}
\end{figure}

% \begin{table}
%     \centering
%     \begin{tabular}{|c|c|} \hline 
%          \textbf{Critic learn rate}& $2 \times 10^{-3}$\\ 
%          \textbf{Actor learn rate}& $5 \times 10^{-3}$\\ 
%          \textbf{Critic/actor gradient threshold}& $1$\\ 
%          \textbf{Discount factor}& $0.999$\\ 
%          \textbf{Mini-batch size}& $64$\\ 
%          \textbf{Noise variance}& $0.1$\\ 
%          \textbf{Noise decay}& $10^{-4}$\\ \hline 
%     \end{tabular}
%     \caption{RL agent hyper-parameters}
%     \label{tab:rlhypers}
% \end{table}

\bibliography{references}

\end{document}